\documentclass{article} 
\usepackage{nips13submit_e,times}
\usepackage{hyperref}
\usepackage{url}
\usepackage{graphicx}
\usepackage{wrapfig}
\graphicspath{ {./images/} }
\usepackage{import}
\usepackage{cite}
\usepackage{subcaption}
\usepackage{amsmath}
\usepackage{amsfonts}

\newcommand{\e}{\mathbb E}

\title{Audio inpainting with generative adversarial network}

\author{
Pirmin Philipp Ebner\\
ETH Zurich
\\
ebnerp@student.ethz.ch
\And
Amr Eltelt\\
ETH Zurich\\
aeltelt@student.ethz.ch
}

\nipsfinalcopy 

\begin{document}

\maketitle

\begin{abstract}
We study the ability of Wasserstein Generative Adversarial Network (WGAN) to generate missing audio content which is, in context, (statistically similar) to the sound and the neighboring borders. We deal with the challenge of audio inpainting long range gaps (500 ms) using WGAN models. We improved the quality of the inpainting part using a new proposed WGAN architecture that uses a short-range and a long-range neighboring borders compared to the classical WGAN model. The performance was compared with two different audio instruments (piano and guitar) and on virtuoso pianists together with a string orchestra. The objective difference grading (ODG) was used to evaluate the performance of both architectures. The proposed model outperforms the classical WGAN model and improves the reconstruction of high-frequency content. Further, we got better results for instruments where the frequency spectrum is mainly in the lower range where small noises are less annoying for human ear and the inpainting part is more perceptible. Finally, we could show that better test results for audio dataset were reached where a particular instrument is accompanist by other instruments if we train the network only on this particular instrument neglecting the other instruments. 

\end{abstract}
\let\thefootnote\relax\footnote{Model: \url{https://github.com/nperraud/gan_audio_inpainting}}
\section{Introduction}
Corrupted audio files, lost information in audio transmission (e.g. voice-over-IP transmission), and audio signals locally contaminated by noise are highly important problems in various audio processing tasks like in music enhancement and restoration. Data loss or corruption in the range of seconds can have various causes like partially damaged physical media. Further, loss of the connection in audio transmission may lead to data loss beyond just a few hundred milliseconds. This has highly unpleasant consequences for a listener, and it is hardly feasible to reconstruct the lost content from local information only.

Larger gaps spanning for seconds could happen in various applications and cases, such as in music enhancement and restoration. Rather unrealistic assumption are often made during short-range inpaining that the signal is stationary \cite{perraudin2018}, which tends not to hold for longer gaps. Long audio inpainting ($>200$ ms) for editing is a very challenging task as long gaps are commonly at the scale of seconds, rendering the algorithms for shorter gaps in vain. Further, the audio features are high dimensional, complex and non-correlated and thus directly applying state-of-the art models in image or video inpaintings tends not to work well \cite{chang2019}.

Despite impressive recent advances with neural networks \cite{van2016wavenet, donahue2018wavegan}, audio recovering remain an important challenge in machine learning. Restoring of lost information in audio has been referred to as waveform substitution \cite{goodman1986}, audio inter/extrapolation \cite{kauppinen2001, etter1996}, or audio inpainting \cite{adler2012}. Reconstruction is usually aimed at providing a coherent and meaningful information while preventing audible artifacts so that the listener remains unaware of any occurred problem. When missing parts have a length no longer than 50ms, sparsity-based techniques can be successful \cite{adler2012, siedenburg2013}. Otherwise, techniques relying on auto-regressive modeling \cite{etter1996}, sinusoidal modeling \cite{lagrange2005, lukin2008} or based on self-content \cite{bahat2015} have been proposed. For greater length, it is useless to try recovering the exact waveform. Fortunately, Generative Adversarial Networks (GANs, \cite{goodfellow2014}) are able to produce signals statistically similar without the need to minimize an L2 loss. Hence they are a very good candidate for the audio inpainting with large gaps problem.

GANs rely on two competing neural networks trained simultaneously in a two-player min-max game: The generator produces new data from samples of a random variable; The discriminator attempts to distinguish between these generated and real data. During the training, the generator's objective is to fool the discriminator, while the discriminator attempts to learn to better classify real and generated (fake) data. The potential advantages of GAN-based approaches is the rapid and straightforward sampling of large amounts of audio. Several GAN approaches have been investigate to fill gaps in audio like WaveGAN \cite{donahue2018wavegan} or TiFGAN \cite{marafioti2019adversarial}. Wavenet \cite{van2016wavenet} is another approach by directly modelling waveforms using a neural network method trained with recordings of real sound.

In the reconstruction of audio signals the type of instrument plays an additional role. The special acoustic impression that a musical instrument makes on us is called the timbre of the instrument. The timbre of an instrument is mainly determined by the additional vibrations in addition to the basic vibrations. The fundamental vibration is the vibration of a string with the greatest amplitude. It determines the pitch. The additional vibrations that are added, which are also called harmonics or overtones, overlap with the fundamental vibration. This superposition of fundamental and harmonics results in a very characteristic vibration pattern for each instrument. We perceive these different vibration patterns as different timbres, also called sound patterns, at one and the same pitch. The number and intensity of the harmonics varies greatly from instrument to instrument. Because of these different harmonics, even instruments of the same type can have different timbres. Having multiple different instrument in the training set, a higher frequency spectrum has to be covered and the generated audio signal is stronger influenced by noise.

The objective of this work is audio inpainting of long gap audio content, in particular the gap addressed is in the range of 500 - 550 ms, depending on the sample rate of the waveform. The focus here is to investigate the use of WGAN in generating audio with global coherence, the impact of architecture improvement on the audio quality, the impact of training with different type of audio instruments with different dataset sizes, and to investigate the possibility of finding a optimal model setting that is generalized rather than overfitted to specific dataset.

In this work we focus on waveform strategies rather than spectrogram or multimodal strategies.
We realise in this task that there is no specific metric that can be used to evaluate the quality of generated audio signal and therefore we depend on human judgement using the objective difference grading technique for evaluation. Furthermore, it is clear that in this unsupervised setting, our goal is not to generate a perfectly matching audio signal given the wide variation of patterns and rhythms in each audio signal trained on.


\section{Methods}

\subsection{Generative Adversarial Network (GAN)}
GANs learn mapping from low-dimensional latent vectors $\textit{\textbf{z}}\in Z$, i.i.d. samples from known prior $P_Z$, to points in the space of natural data $\mathcal{X}$. A generator $G : \mathcal{Z} \mapsto \mathcal{X}$ is pitted against a discriminator $D : \mathcal{X} \mapsto \left[0,1\right]$ in a two-player minimax game \cite{goodfellow2014}. $G$ is trained to minimize the following value function, while $D$ is trained to maximize it:

\begin{equation}
    V(D,G) = \e_{\textit{\textbf{x}}\sim P_X} \left[\log D(\textit{\textbf{x}})\right] + \e_{\textit{\textbf{z}}\sim P_Z} \left[\log (1-D(G(\textit{\textbf{x}})))\right] \\
    \label{eq:valfct}
\end{equation}

$D$ is trained to determine if an example is real or fake, and $G$ is trained to fool the discriminator into thinking its output is real. Equation \ref{eq:valfct} equates to minimizing the Jensen-Shannon lower bound between $P_X$, the data distribution, and $P_G$, the implicit distribution of the generator when $\textit{\textbf{z}}\sim P_Z$. The Jensen-Shannon divergence measure the similarity between two probability distribution, bounded by $\left[0,1\right]$, and is symmetric:

\begin{equation}
    D_{JS}(P_X\| P_Z) = \frac{1}{2}D_{KL}\left(P_X\| \frac{P_X+P_Z}{2}\right) +\frac{1}{2}D_{KL}\left(P_Z\| \frac{P_X+P_Z}{2}\right)
\end{equation}

where $D_{KL}(P_X\| P_Z)$ is the Kullback-Leibler divergence measuring how the probability distribution $P_X$ diverges from the second expected probability distribution $P_Z$:

\begin{equation}
    D_{KL}(P_X\| P_Z)=\int_x P_X(x)\log \frac{P_X(x)}{P_Z(x)}dx
\end{equation}

In this formulation, GAN are notoriously difficult to train, and prone to catastrophic failure cases, with mode collapse and oscillations a common problem. Instead of producing samples sufficiently representing the true data distribution, the generator maps the entire latent space to a limited subset of the real data space. The generator does not 'converge' to a stationary distribution and \textit{mode collapse} occurs \cite{che2017, srivastava2017}. Further, GANs have \textit{catastrophic forgetting} tendency, which leads to the discriminator losing the ability to remember synthesized samples from previous instantiations of the generator \cite{liang2018}.

\subsection{Wasserstein Generative Adversarial Network (WGAN)}
It is an extension of the GAN that seeks an alternate way of training the generator model to better approximate the distribution of data observed in a given training dataset.

Instead of using a discriminator to classify or predict the probability of generated images as being real or fake, the WGAN changes or replaces the discriminator model with a critic that scores the realness or fakeness of a given sample. The critic, however, can’t saturate, and converges to a linear function that gives remarkably clean gradients everywhere. 

This change is motivated by a theoretical argument that training the generator should seek a minimization of the distance between the distribution of the data observed in the training dataset and the distribution observed in generated examples.

Instead of Jensen-Shannon divergence, the smoother Wasserstein-1 distance between generated and data distribution is minimized \cite{arjovsky2017}:

\begin{equation}
    W(P_X,P_G) = \underset{\| f\| _L \leq 1}{\textrm{sup}} \e_{\textit{\textbf{x}}\sim P_X} \left[f(x)\right] - \e_{\textit{\textbf{x}}\sim P_G} \left[f(x)\right]
\end{equation}

where $\| f \| _L \leq 1 : \mathcal{X} \mapsto \mathbb{R}$ is the family of functions that are 1-Lipschitz. The Wasserstein distance is minimized for the following value function

\begin{equation}
    V_\textrm{WGAN}(D_w,G) = \e_{\textit{\textbf{x}}\sim P_X}\left[D_w(\textit{\textbf{x}})\right] - \e_{\textit{\textbf{z}}\sim P_Z}\left[D_w(G(\textit{\textbf{z}}))\right]
\end{equation}

\begin{figure}[h]
    \includegraphics[scale=0.36]{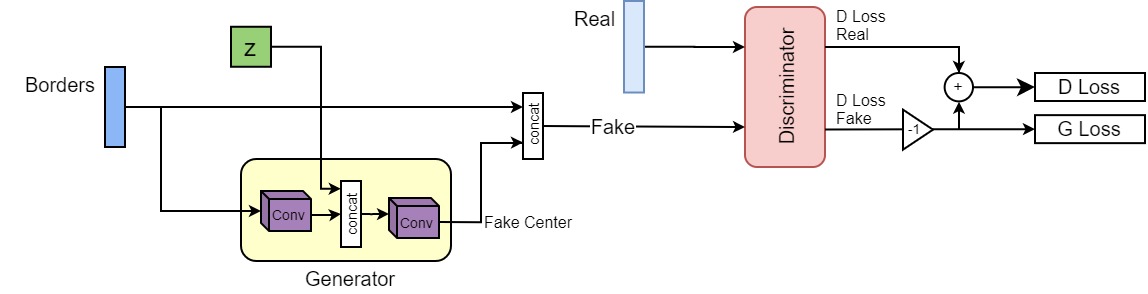}
    \centering
    \caption{Wasserstein GAN (WGAN).}
    \label{fig:wgan}
\end{figure}

$D_w : \mathcal{X} \mapsto \mathbb{R}$ is not trained to identify examples as real or fake, but instead is trained as a function that assists in computing the Wasserstein distance (Figure \ref{fig:wgan}).

The benefit of the WGAN is that the training process is more stable. Perhaps most importantly, the loss of the discriminator appears to relate to the quality of images created by the generator.

\section{Dual Discriminator WGAN}
To tackle the problem of long period audio inpainting we propose in this work an approach for segmenting the raw audio waveform into borders neighboring the missing audio content and furthermore applying a new model architecture which includes the Wasserstein GAN but with an additional discriminator component applied. This new architecture is designed to accommodate the new borders segmentation. 

\subsection{Long and Short Length Borders}

\begin{wrapfigure}{r}{0.37\textwidth}
    \includegraphics[scale=0.7]{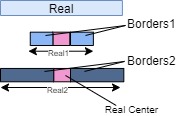}
    \centering
    \caption{Long and short neighbouring borders segmented from the real audio.}
    \label{fig:borders}
\end{wrapfigure}

The approach applied here is the extraction of the short-range borders and the long-range borders of the missing audio content. The strategy behind this method is based on the idea that close neighbor audio have proven success to inpaint short period missing audio, however it fails for long periods. A larger range of neighboring audio can tell us something about the long missing audio content. By combine both overlapped borders we can involve more correlated information to the missing data and use it to train our model.


As shown in the Figure \ref{fig:borders}, the batch of real audio is split into segments around the missing content. The missing audio is set to around 500 ms (Depending on the sampling rate of the audio signal). Borders1 (Shorter borders) has a length of less than 1 s each side while Borders2 has a length of 2.5 s each side.

\subsection{Model Architecture}
One of the issues with GAN is the mode collapse which is formulated by the problem of the minimax optimization. Recent attempts have been made to solve the mode collapsing problem by improving the training \cite{Hoang2017MultiGenerator}. The idea of applying a dual discriminator GAN (D2GAN) as a method to tackle the mode collapse has been discussed in \cite{nguyen2017dual}.

\begin{figure}[h]
    \includegraphics[scale=0.36]{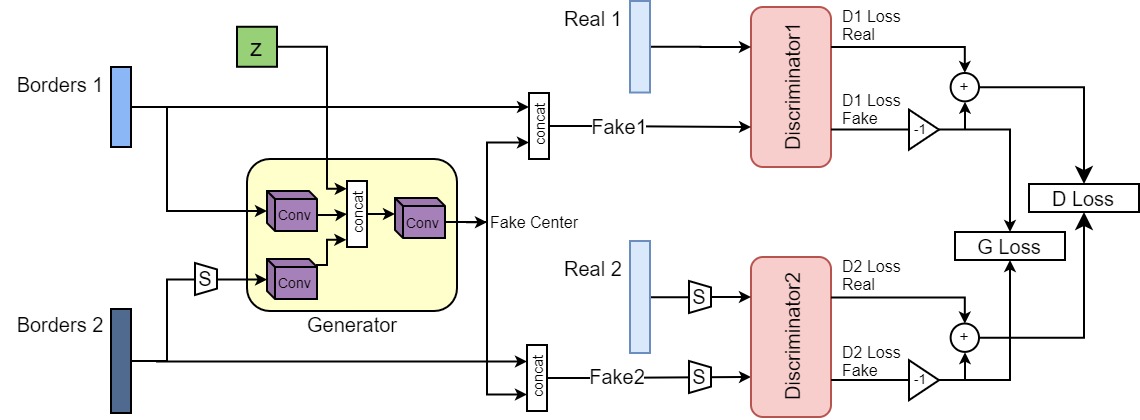}
    \centering
    \caption{Dual Discriminator Wasserstein GAN (D2WGSN).}
    \label{fig:extendedmodel}
\end{figure}

We propose here a new approach motivated by \cite{nguyen2017dual} and based on the Wasserstein loss function. We term our proposed model as Dual Discriminator Wasserstein GAN (D2WGAN).

As shown in Figure \ref{fig:extendedmodel}, the generator inputs includes Borders 1 and 2 along with the normally distributed latent variable. The generator use a convolution operation for each input then concatenated all with the latent variable as an input to another convolution operation. Both discriminators are identical and includes components for convolution operations as well. The code of the generator and discriminator components are referenced from the Tensorflow implementation \cite{codegan} and has been adapted and integrated to the new model setup.

\subsection{Loss Function}
\begin{figure}[h!]
  \centering
  \begin{subfigure}[b]{0.4\linewidth}
    \includegraphics[width=\linewidth]{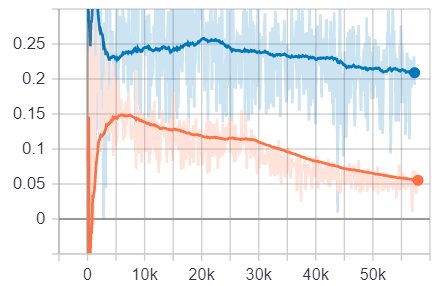}
    \caption{Total discriminator loss}
  \end{subfigure}
  \begin{subfigure}[b]{0.4\linewidth}
    \includegraphics[width=\linewidth]{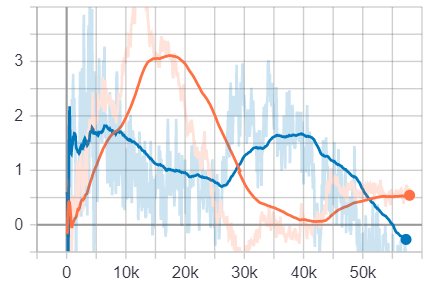}
    \caption{Generator loss}
  \end{subfigure}
  \caption{D2WGAN training with comparison between two different audio datasets. PIANO (Orange) and SOLO (Blue).}
  \label{fig:loss}
\end{figure}

The network is trained on optimizing the Wasserstein loss. In this architecture we minimize the total loss of both discriminators. In addition, we minimize the generator loss which is the total of the negation of discriminators fake losses, see Figure \ref{fig:extendedmodel}. Figure \ref{fig:loss}, shows the performance of training the total D-loss and total G-loss for the PIANO and SOLO dataset (more details about the dataset are given in section \ref{subsec:Dataset}) . 

\section{Evaluation}
The main objective of the evaluation was to investigate our networks's ability to adapt to different audio signals. The evaluation is based on a comparison of the inpainting quality by means of objective difference grades (ODGs \cite{recommendatioin2001}) and looking at the magnitude spectrograms.

We considered two classes of audio signal: instrument sounds, mainly piano and guitar and virtuoso pianists together with a string orchestra. The two networks were trained on the targeted signal class, with an gap size of 500 ms. Reconstruction was evaluated on the testing data for 500 ms gaps.

\subsection{Parameters}
The sampling rate was 16 kHz and audio segments with a duration of 6.5 s, corresponding to $L=53248$ samples for both models were considered. Each segment was separated in a gap of 500 ms corresponding to $L_g=4096$ of the central part of a segment and the context of twice of 3.0 s, corresponding to $L_c=24576$ samples.

\subsection{Dataset}\label{subsec:Dataset}
We considered instrument sounds including piano, acoustic guitar and piano together with a string orchestra using three different dataset. The respective networks were trained on the targeted signal, with an gap size of 0.5 s. Reconstruction was evaluated on the trained signal class. The datasets were splitted into training, validation, and testing sets before segmenting them. The statistics of the resulting sets are presented in Table \ref{tab:maestro} for the following datasets:

\begin{table}[ht]
    \centering
    \begin{tabular}{c|c|c|c|c|c}
        \hline
         &      & Total & Training & Validation & Testing \\
        \hline
        PIANO   & Files     & 19 & 15 & 2 & 2 \\
                & Duration  & 26.7 min & 20.4 min & 3.3 min & 3.0 min \\
        \hline
        SOLO    & Files     & 48 & 38 & 5 & 5 \\
                & Duration  & 144 min & 114 min & 14.3 min & 15.7 min\\
        \hline
        MAESTRO & Files     & 1282 & 967 & 137 & 178 \\
                & Duration  & 201.2 h & 161.3 h & 20.5 h & 19.4 h \\
        \hline
    \end{tabular}
    \caption{Dataset information}
    \label{tab:maestro}
\end{table}

\begin{itemize}
    \item \textbf{PIANO}: The PIANO dataset is the smallest dataset to check the performance of our two networks. It contains only piano samples from Johann Sebastian Bach and the uncompressed audio is of CD quality (49 kHz 16-bit PCM stereo). The dataset can be downloaded here \url{http://deepyeti.ucsd.edu/cdonahue/wavegan/data/mancini_piano.tar.gz}. An overview about the dataset is given in Table \ref{tab:maestro}.
    \item \textbf{SOLO}: The SOLO dataset is a collection of different instruments recording including accordion, acoustic guitar, cello, flute, saxophone, trumpet, violin, and xylophone. The dataset is available on Kaggle (\url{https://www.kaggle.com/zhousl16/solo-audio}). The uncompressed audio is of CD quality (44.1 kHz 16-bit PCM mono) and splitted in train/validation/test. In our work only the acoustic guitar dataset was used to train and test our models. Table \ref{tab:maestro} gives an overview about the data.
    \item \textbf{MAESTRO}: The biggest used dataset for this work is the MAESTRO ('MIDI and Audio Edited for Synchronous TRacks and Organization') dataset based on recordings from the International Piano-e-Competition (\url{http://piano-e-competition.com/}). It is a piano performance competition where virtuoso pianists perform on Yamaha claviers together with a string orchestra. The dataset is available at \url{https://magenta.tensorflow.org/maestro-wave2midi2wave}. In particular, MAESTRO contains over 201 hours of paired audio and MIDI recordings from nine years of the competition, significantly more data than similar datasets. Audio and MIDI files are aligned with $\approx 3 \textrm{ ms}$ accuracy and sliced to individual musical pieces. Uncompressed audio is of CD quality or higher (44.1 - 48 kHz 16-bit PCM stereo). The data are splitted in train/validation/test so that the same composition, even if performed by multiple contestants, does not appear in multiple subsets. Repertoire is mostly classical, including composers from the $17^\textrm{th}$ to early $20^\textrm{th}$ century. Table \ref{tab:maestro} gives an overview about the data.
\end{itemize}

\subsection{Evaluation metrics}
Using the signal-to-noise ratios (SNRs) applied to the time-domain waveforms and magnitude spectrograms is not an appropriate indicator to evaluate the model because WGAN is not optimizing the L2-loss. Therefore, we conducted a user study evaluating the inpainting quality by means of objective difference grades (ODG \cite{recommendatioin2001}). It correspond to the subjective difference grade used in human-based audio test and is derived from the perceptual evaluation of audio quality (PEAQ, \cite{recommendatioin2001}). ODG range from 0 to -4 with the interpretation shown in Tab. \ref{tab:odg}.

We generated 50 samples of 6.5 s length for each dataset and model, in total 150 samples per model, evaluate them by two independent person without knowing the source of the samples and evaluate using ODG.

\begin{table}[ht]
    \centering
    \begin{tabular}{l|l}
        \hline
        ODG & Impairment \\
        \hline
        0   & Imperceptible \\
        -1  & Perceptible, but not annoying \\ 
        -2  & Slightly annoying \\
        -3  & Annoying \\
        -4  & Very annoying \\
        \hline
    \end{tabular}
    \caption{Interpretation of ODGs.}
    \label{tab:odg}
\end{table}

\subsection{Generator}
A data loader generator is used for the MAESTRO dataset because it is too big for the memory. The generator is needed to feed the model with the dataset. It allow us to lazily evaluate data. This concept of lazy evaluation is useful because it lets us generate values in an efficient manner by yielding only chunks of data at a time instead of the whole thing at once.

\subsection{Audio signal preprocessing}
The PIANO, SOLO and MAESTRO signals are preprocessed before being loaded. In addition to normalizing the data and establishing a mono signal, the audios are filtered using FIR low pass filter using the window method. The filter cutoff frequency is set lower than Nyquist frequency. This ensures that we filter out any high frequency noise signals higher than the Nyquist, thus avoid any aliasing effect. Furthermore, the audio signals are then downsampled by 3.

The PIANO and MAESTRO sample rate is 48 kHz (Nyquist Freq. = 24 kHz), the cutoff frequency is set to 13.2 kHz, and the final signal is downsampled to 16kHz. The SOLO sample rate is 44.1 kHz, the cutoff frequency is set to 11.025 kHz, and downsampled to 14.7 kHz.

\subsection{Training}
Both existing and new implemented WGAN networks were trained for different instrument and music dataset, resulting in six trained networks. Each training started with the learning rate of $10^{-4}$ and using the Adam-optimizer for both the generator and discriminator. Batch-size of 64 was used, every 100 steps, the training progress was monitored, and every 1000 steps the model was saved. The PIANO and SOLO dataset was trained for 40000 and 56300 steps and the MAESTRO dataset for 73000 and 83000 steps for the classic WGAN and D2WGAN model, respectively. 

\begin{figure}[h]
    \includegraphics[width=0.6\textwidth]{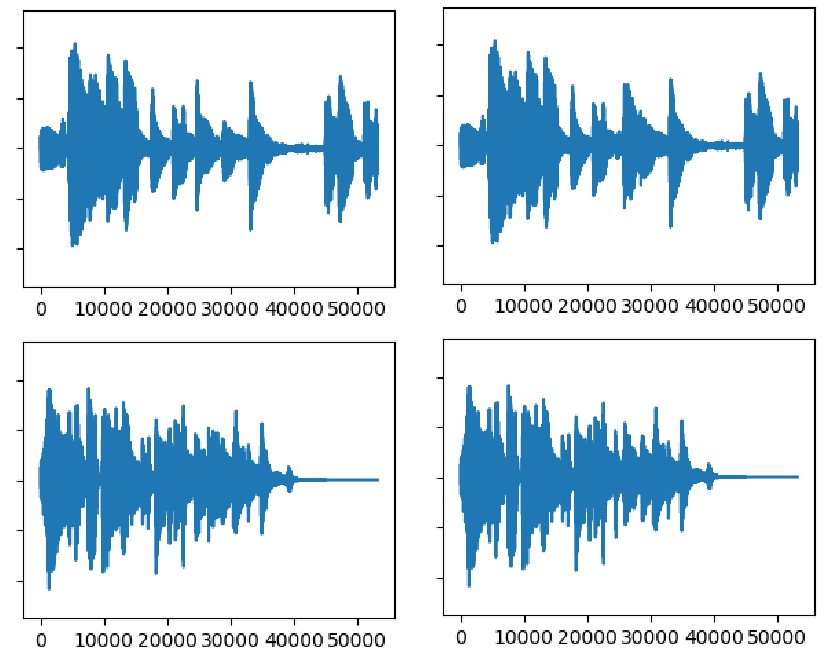}
    \centering
    \caption{A comparison of samples between the real signal images (left) and the D2WGAN fake images (right)}
    \label{fig:realfakeimages}
\end{figure}

From several experiments it has been noticed that the number of training iterations have direct impact on improving the quality of generated fake audio. Figure \ref{fig:realfakeimages}, shows two examples of the improvement of similarity between real and D2WGAN fake audio images along the training iterations.Therefore, it is essential to reduce training time and hence gain more iterations. With the D2WGAN setup the excepted training time required would be much higher than the classic WGAN, given that we use two discriminators and two set of borders for the generator. Therefore, to reduce training time the longer borders (Borders 2) and the corresponding longer real (Real 2) are downsampled by 4, before entering the generator and discriminator. Refer to figure \ref{fig:extendedmodel}.

\section{Results and discussion}
We analysed the performance of the WGAN and the D2WGAN architecture using all dataset. Further, a comparison between the datasets was performed and the effect of training steps and datasets on the D2WGAN model was analysed.

\subsection{Effect of the network type}
\begin{figure}[h!]
  \centering
  \begin{subfigure}[b]{0.45\linewidth}
    \centering
    \includegraphics[width=\linewidth]{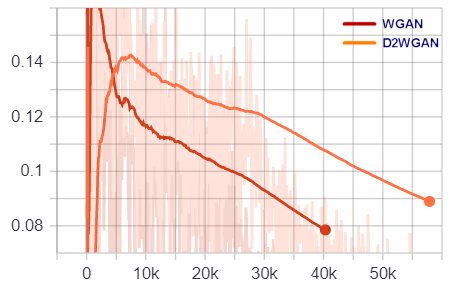}
    \caption{PIANO discriminator D-loss}
  \end{subfigure}
  \begin{subfigure}[b]{0.45\linewidth}
    \centering
    \includegraphics[width=\linewidth]{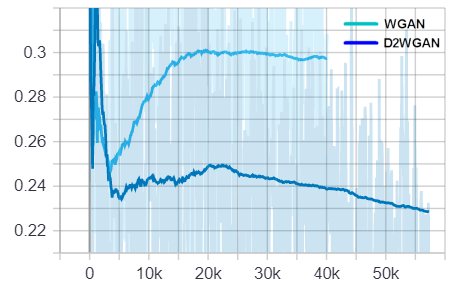}
    \caption{SOLO discriminator D-loss}
  \end{subfigure}
  \begin{subfigure}[b]{0.45\linewidth}
    \centering
    \includegraphics[width=\linewidth]{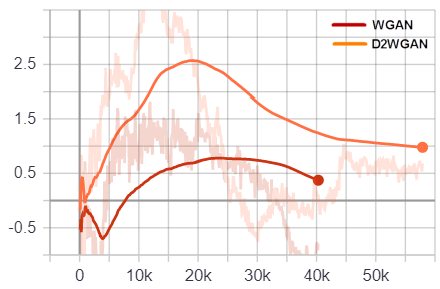}
    \caption{PIANO generator G-loss}
  \end{subfigure}
  \begin{subfigure}[b]{0.45\linewidth}
    \centering
    \includegraphics[width=\linewidth]{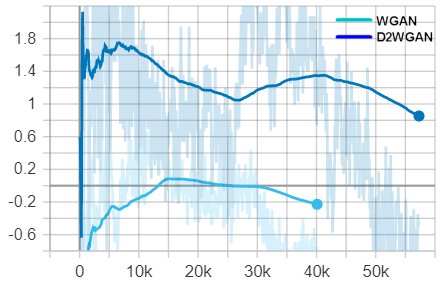}
    \caption{SOLO generator G-loss}
  \end{subfigure}
  \caption{The convergence of both the discriminator loss and generator loss training on the WGAN and D2WGAN models for the PIANO and SOLO datasets.}
  \label{fig:entendvsbasic52}
\end{figure}

In order to compare between the two network types, each model was trained on the three datasets. The training performance of each model gives insights on the expected contribution of the new model. From Figure \ref{fig:entendvsbasic52} (a) and (c), we can see in case of the PIANO dataset both models are converging the loss, however the WGAN G-loss is outperforming that of the D2WGAN as shown in (c). While as shown in (b) and (d), the D2WGAN model on the SOLO dataset is highly outperforming the WGAN model in terms of D-loss and G-loss convergence.

Moreover, each network type was tested on all three datasets by generating fake audios using a separate test set and reconstructing the full length audio signal and therefore we are able to evaluate audio quality using ODG. Table \ref{tab:odgs} shows the ODGs of those predictions. 
On the SOLO dataset its observed that there is an improvement in the audio signal using the D2WGAN network. The mean ODG score is -2.46 while for the WGAN it scored -2.54. Furthermore, the standard deviation values are in favor of the D2WGAN model. On the other hand, with the PIANO dataset the D2WGAN had a slight improvement on the WGAN with a mean of -2.54 for D2WGAN compared to -2.56 of WGAN. This reflected also by the training performance of the WGAN generator as shown in \ref{fig:entendvsbasic52} (c).
Furthermore, the MAESTRO data set performance was aligned with the SOLO scores, where we also see the D2WGAN audio output quality much improved compared of that of WGAN.
On average the D2WGAN model with the extended borders scheme gave better results with an overall mean ODG of -2.67 compared to -2.77 of the WGAN.

\subsection{Comparison between the datasets}

\begin{table}[ht]
    \centering
    \begin{tabular}{c | c c c | c c c}
         \hline
         \\[-0.8em]
         & \multicolumn{3}{c|}{WGAN} & \multicolumn{3}{c}{D2WGAN} \\
         & PIANO & SOLO & MAESTRO & PIANO & SOLO & MAESTRO \\
        \hline
        \\[-0.8em]
        ODG: \ 0     & 0 & 0 & 0 & 0 & 2 & 0 \\
        ODG: -1    & 1 & 6 & 0 & 0 & 4 & 1 \\
        ODG: -2    & 20 & 16 & 5 & 23 & 16 & 13 \\
        ODG: -3    & 29 & 23 & 29 & 27 & 25 & 21 \\
        ODG: -4    & 0 & 5 & 16 & 0 & 3 & 15 \\
        \hline
        \\[-0.8em]
        Mean ODG    & -2.56 & -2.54 & -3.2 & -2.54 & -2.46 & -3.0 \\
        Std ODG     & 0.53 & 0.83 & 0.61 & 0.87 & 0.5 & 0.81 \\
        \hline
        \\[-0.8em]
        Overall mean ODG & \multicolumn{3}{c|}{-2.77} & \multicolumn{3}{c}{-2.67} \\
        Overall Std ODG & \multicolumn{3}{c|}{0.66} & \multicolumn{3}{c}{0.73} \\
        \hline
    \end{tabular}
    \caption{The counted ODGs and the corresponding mean and standard deviation (Std) of reconstructions of 0.5 s gaps for the classic WGAN and D2WGAN model.}
    \label{tab:odgs}
\end{table}

Table \ref{tab:odgs} provides counted ODGs and the corresponding mean and standard deviation for each datasets and model. The D2WGAN slightly outperformed the classic WGAN model for each dataset and has also a slightly higher standard deviation. Looking at the counted ODGs it is obvious that the D2WGAN model has higher ODG values confirming the higher standard deviation. Both WGAN network seem to better inpaint the SOLO dataset compared to the PIANO dataset. Further, the two small dataset PIANO and SOLO have better inpainting quality compared to the huge MAESTRO dataset. Possible reasons are:
\begin{itemize}
    \item The SOLO dataset, containing only guitar sounds, offers much more inpainting techniques and sound variations than the PIANO dataset, containing only piano sounds. A guitar has twenty or twenty-four frets per string. This means that there are a total of 144 notes. On the piano it is only 88 tones. Therefore, a guitar covers a wider range of notes and small noise or interruptions can be less recognized.
    \item A guitar only covers four octaves where the tones are softer, while a piano covers seven octaves and the tones are mainly harder. This means that a piano has also an impact in the higher frequency domain and therefore, the training complexity increases.
    \item The MAESTRO dataset, a combination of piano together with a string orchestra, is much harder to train. On one side it is a huge dataset of 201 hours of music and there is not only one single instrument playing. Due to the heterogeneity of the ground truth audio quality the training complexity increases. This sometimes leads to notice “timbral shifts” in the reconstructed audio files.
\end{itemize}

Important point, compared to images reconstruction using WGAN models, small noise ($< 1 \%$) in audio signal can have a huge impact on the quality of the sound. Especially, the human ear has different sensitivities to different frequencies. Our ears being more sensitive to upper frequencies for example, a pure-tone of 1000 Hz with intensity level of 40 dB would impress the human ear as being louder than a pure-tone of 80 Hz with 50 dB, and a 1000 Hz tone at 70 dB would give the same subjective impression of loudness as a 50 Hz tone at 85 dB. If there is a corruption in the inpainted signal there is a big change of the amplitude of the signal compared with the original (the signal is supposed to be smooth). Thus, a signal of low frequency really do not represent a great change in the whole signal because it represent a small portion of the information if we take a fix period of analysis. Therefore, lower frequency signal represents information at a lower rate (needs lesser samples to reconstruct it) whereas a high frequency signal represents information at a faster rate (and hence needs more samples while reconstruction). Noise could affect both the signals in a similar way of loosing/corrupting some information (sample values) from those signals. If we increase the frequency of the signal, the change produced by noise represents a higher proportion of the whole signal and the corruption is higher compared with the low frequency signal. Thus, it needs more samples to represent higher frequency signals than for the lower frequency signal. This could be the explanation why SOLO is better than PIANO and MAESTRO. 

\begin{figure}[ht]
    \includegraphics[scale=0.124]{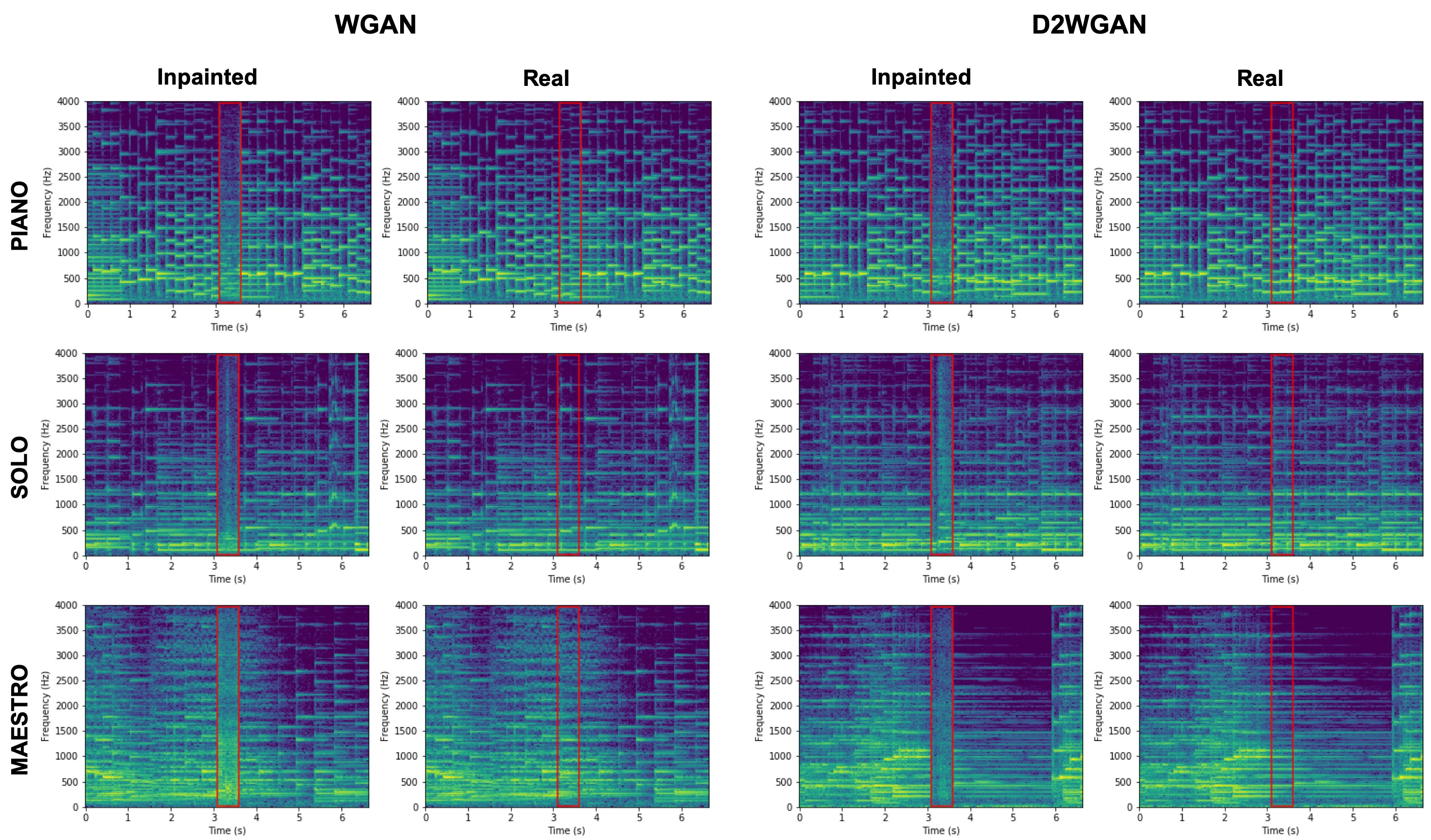}
    \centering
    \caption{Magnitude spectrograms (in dB) of an exemplary signal reconstruction. Left: inpainted and real signal of the classic WGAN model; Right: inpainted and real signal of the D2WGAN model. From top to bottom each dataset is plotted. The gap was the area between the two red lines. The ODG quality of the datasets is between -1 and -2.}
    \label{fig:spectrogram}
\end{figure}

Figure \ref{fig:spectrogram} shows the magnitude spectrogram, visualizing the change of a nonstationary signal’s frequency content over time, of the real and reconstructed sound for all three dataset and both model. The first two columns show the inpainted and real data of the classic WGAN and the last two columns the inpainted and real data of the D2WGAN model. Obviously, PIANO shows strong developed magnitude spectrograms pattern at high frequency compared to SOLO and MAESTRO. MAESTRO has also high intensity at higher frequency but they are more smoothed over the frequency domain and more irregular in times due to the background orchestra. The worse performance using the MAESTRO dataset is due to the higher irregularity at higher frequency and interaction of the different instruments.

\subsection{Effect of training steps and datasets on the D2WGAN model}\label{subsec:Effect of steps}
We investigated the effect on what happens if we train the model on one dataset but tested it on another one. Further, we improved our ODG performance by training our network longer. Table \ref{tab:odgs_spec} shows the counted ODGs and the corresponding mean and standard deviation using the D2WGAN model. The D2WGAN model was trained on the PIANO dataset for 140000 steps and tested both on the PIANO and MAESTRO dataset. The D2WGAN model showed on both dataset a huge improvement of the ODGs. The results indicate that we didn't overfit the D2WGAN model and we got excellent performance.

\begin{table}[ht]
    \centering
    \begin{tabular}{c | c c}
         \hline
         & \multicolumn{2}{c}{D2WGAN - PIANO}\\
         & PIANO & MAESTRO \\
        \hline
        ODG: \ 0   & 0 & 3 \\
        ODG: -1    & 27 & 35 \\
        ODG: -2    & 22 & 12 \\
        ODG: -3    & 1 & 0 \\
        ODG: -4    & 0 & 0 \\
        \hline
        Mean ODG    & -1.48 & -1.18 \\
        Std ODG     & 0.54 & 0.52 \\
        \hline
        Overall mean ODG & \multicolumn{2}{c}{1.33} \\
        Overall Std ODG & \multicolumn{2}{c}{0.53}  \\
        \hline
    \end{tabular}
    \caption{The counted ODGs and the corresponding mean and standard deviation (Std) of reconstructions of 0.5s gaps for the D2WGAN model trained on the PIANO dataset and tested with PIANO and MAESTRO dataset.}
    \label{tab:odgs_spec}
\end{table}

Interestingly, the trained D2WGAN model slightly outperformed on the MAESTRO dataset compared to the PIANO dataset although PIANO dataset was used for training. One explaination could be that the model interpreted the orchestra mainly as background noise. Thus, small noise in the inpainting part will be stronger smoothed and therefore the impairment is more perceptible but not annoying. Apparently, training the MAESTRO dataset mainly on piano and suppress the orchestra can improve the performance of the network.

\section{Conclusion and outlook}
In this project we analysed long gap (500 - 550 ms) audio content inpainting using two kind of GANs: the classical Wasserstein GAN (WGAN) and the Dual Discriminator WGAN (D2WGAN) models. The D2WGAN model is a new proposed architecture improvement to inpaint the missing audio content using a short-range and a long-range borders. Both models were trained with different type of audio instruments (piano and guitar) and virtuoso pianists together with a string orchestra to investigate the possibility of finding an optimal model that is generalized rather than overfitting to specific dataset. The testing was evaluated by human judgement using the objective difference grading (ODG).

The new proposed D2WGAN model slightly outperforms the classic WGAN model for all three different dataset. We observed an improvement in the reconstruction of high-frequency content leading to better ODG scores. Especially, a better improvement was observed for the guitar and virtuoso pianists plays together with a string orchestra dataset. 
This can be associated with the fact that the long range borders combined with short range borders can potential provide information in correlation with the gap content. The idea of two discriminators take advantage of the extend border. 
Furthermore, the poor results for the PIANO can be illustrated by the fact that the PIANO dataset lack audible variation in sound. This can be shown by the overall ODG results of the PIANO compared to SOLO and MAESTRO. The selected training dataset has a direct impact on the audio inpainting application. As we aim for generalized solution, we need to have a dataset with a generalized audio content. 

Generally, better results can be expected for instruments where the frequency spectrum is in the lower range and the higher frequencies are less affected and for sound of more dynamic instruments. This is the case for guitar or string instruments. Therefore, small noise are less annoying for the human ear and the inpainting part is more perceptible.

We also showed that by increasing the training steps up to 140k, the D2WGAN has significantly improved performance without overfitting as shown in table \ref{tab:odgs_spec} by comparing the results of testing two different datasets, which also means there is still room for improvement. More interestingly, we showed that better results can be achieved for audio dataset where a particular instrument is accompanist by other instruments if we train the network only on this particular instrument and neglecting the other instruments. For example, training on PIANO and testing on MAESTRO.

In the future to improve the model and the training performance we suggest to do following: (1) Playing around with the boarder length to see whether providing more information to the model helps to improve the performance. Furthermore, by changing the long and small range borders do we see improvements? (2) So far we have fixed the generator and discriminator architectures. It would be useful to understand the impact of adding more layers to the convolution operation or changing filters sizes. (3) Examine further the loss function. Instead of adding the loss of both discriminators, we investigate the impact of factorizing the D-losses giving more weight for one discriminator on the other based on their contributions. (4) Explore Multimodal strategies where we combine waveform and spectrogram images. (5) Investigate the impact of the frequency range on the model performance. Do we get better results for audio sounds having lower frequency signals? (6) Investigate more on our observed result that better results are really expected for audio dataset where a particular instrument is accompanist by other instruments if we train the network only on this particular instrument and neglecting the other instruments. How big is the impact of different kind of accompanying instruments? (7) Training on a wide range of music instruments. For example, the SOLO dataset includes other instruments besides guitar.

\section*{Acknowledgement}
This work is done in collaboration with the Swiss Data Science Center (SDSC) in Zurich, Switzerland. The authors would like to thank SDSC for their support and in particularly Nathanaël Perraudin for his constructive and insightful inputs and for helping to manage the project.

\bibstyle{ieeetr}
\bibliographystyle{ieeetr}
\bibliography{report}

\end{document}